\documentstyle[emulateapj,apjfonts,epsfig]{article}

\slugcomment{}

\begin{document}

\title{Probing the precession of the inner accretion disk in Cygnus X-1}

\author{Diego F. Torres\footnote{Lawrence Livermore National
Laboratory, 7000 East Ave., L-413, Livermore, CA 94550, USA. E-mail:
dtorres@igpp.ucllnl.org}, Gustavo E. Romero\footnote{Instituto
Argentino de Radioastronom\'{\i}a (IAR), C.C.\ 5, 1894 Villa Elisa,
Argentina. E-mail: romero@irma.iar.unlp.edu.ar}, Xavier
Barcons\footnote{Instituto de F\'{\i}sica de Cantabria (CSIC-UC),
39005 Santander, Spain. E-mail: barcons@ifca.unican.es}, Youjun
Lu\footnote{Center for Astrophysics, University of Science \&
Technology of China, Hefei, Anhui 230026, P. R. China. E-mail:
lyj@astron.berkeley.edu } }

\begin{abstract}

We show that changes in the orientation of the inner accretion disk
of Cygnus X-1 affect the shape of the broad Fe K$\alpha$ emission
line emitted from this object, in such a way that eV-level spectral
resolution observations  (such as those that will be carried out by
the {\it ASTRO-E2} satellite) can be used to analyze the dynamics of
the disk. We here present a potential  diagnostic tool, supported by
numerical simulations, by which a few observations of Cygnus X-1,
separated in time, can determine whether its accretion disk actually
precesses, and if so, determine its period and precession angle.
This approach could also be used for similar studies in other
microquasar systems.

\end{abstract}


\section{Introduction}

Cygnus X-1 is a 5.6-day X-ray binary that harbors the best studied
black-hole candidate in the Galaxy. The mass of the accreting
compact object has been estimated as $\sim 10.1$ M$_{\odot}$ and the
donor star is classified as an O9.7 Iab  supergiant of $\sim 17.8$
M$_{\odot}$ (Herrero et al. 1995). The system is located at $\sim 2$
kpc (e.g. Gierli\'nski et al. 1999).
The black hole accretes through the wind of the companion star. Most
of the time, the X-ray source is in the so-called low/hard state,
characterized by a relatively weak blackbody component peaking at a
few keV plus a strong hard power-law of photon index $\sim1.6$. A
nonthermal radio jet has been observed in this state (Stirling et
al. 2001), extending up to $\sim 15$ mas. The jet seems to form an
average angle with the line of sight of $\sim 30^{\circ}$ (Fender
2001) and it has been suggested that it might be precessing
(Stirling et al. 2001, Romero et al. 2002). Occasionally, a
transition to a high/soft state can occur. In this state, most of
the radiated energy is concentrated in the blackbody, while the
power-law component becomes softer, with an index of $\sim 2.8$ and
no jet has been observed.

The usual interpretation of the X-ray behavior of the source is that
the blackbody component originates in a cold, optically thick
accretion disk, whereas the power-law component is produced in an
optically thin hot corona by thermal Comptonization of disk photons
(Poutanen et al. 1997, Dove et al. 1997, Esin et al. 1997, 1998).
The hot corona fills the inner few tens of gravitational radii
around the compact object and the accretion disk would penetrate
only marginally into the coronal region. In the low/hard state the
thermal X-ray luminosity is dominated by the corona, with typical
luminosities of $\sim$ a few times $10^{37}$ erg s$^{-1}$. During
the transition to the high/soft state the corona is likely ejected
as the accretion disk approaches to the black hole (Fender et al.
2004, also Esin et al. 1997, 1998). Most of the energy is then
dissipated by the disk, until the inner part of it dominates the
radiation again, and the cycle starts again.

In the low/hard state, the disk is illuminated by hard photons from
the corona resulting in the production of an Fe K$\alpha$ line and a
Compton reflection feature. The first detection of the line was made
by Barr et al. (1985) with $EXOSAT$. They reported a broad (FWHM
$\sim1.2$ keV) emission line at $\sim 6.2$ keV with an equivalent
width of $\sim120$ eV. Kitamoto et al. (1990) obtained a $Tenma$
GSPC spectrum which was consistent with a narrow emission line at
$\sim 6.5$ keV with an equivalent width of $60-80$ eV. A Compton
reflection feature was then found above 20 keV (see Tanaka 1991 and
references therein). The $Ginga$ spectrum in the $2-30$ keV range
can be fitted quite well by the sum of a power-law with index
$\sim1.7$, a reflection component and a narrow Fe emission line at
$6.4$ keV with an equivalent width of $\sim 60$ eV (Tanaka 1991).
Subsequent $ASCA$ observations confirmed these results but
restricting the width to $10-30$ eV (Ebisawa et al. 1996). A broad
edge at $E>7$ keV was also reported.  A detailed historical
  account of attempts to detect an Fe line in Cygnus X-1 and the
  intrinsic difficulties it entails, can be found in Reynolds \& Nowak (2003).
Recently, {\it Chandra} observed Cygnus X-1 with the High Energy
Transmission Grating Spectrometer in an intermediate X-ray state
(Miller et al. 2002). The narrow Fe line was detected at $E=6.415\pm
0.007$ keV with an equivalent width of $W=16^{+3}_{-2}$ eV, along
with a broad line at $E=5.82\pm 0.07$ keV with $W=140^{+70}_{-40}$
eV. A smeared edge was also detected at $7.3\pm 0.2$ keV. Miller et
al. (2002) interpret these results in terms of an accretion disk
with irradiation of the inner disk producing the broad Fe K$\alpha$
emission line and edge, and irradiation of the outer disk producing
the narrow line. The broad line is shaped by Doppler and
gravitational effects and, to a lesser extent, by Compton
reflection.

For different spectral states, the different disk structure may
change the Fe K$\alpha$ line. For example, if the disk is truncated
at a much larger radius rather than the innermost stable orbit as
suggested for the hard state, then the width of the Fe K$\alpha$
line may significantly decrease. In any case, the variation due to
disk precession, if that were observable, and that due to the
accretion mode (disk structure), will have a different temporal
signature (periodicity), what would make them easy to distinguish.

In this paper, we show that changes in the orientation of the inner
accretion disk of Cygnus X-1 would affect the shape of the broad Fe
K$\alpha$ emission line in a periodic way. Under the
  assumption that the X-ray spectrum of Cyg X-1 is not substantially
  more complex than what has already been found in the {\it Chandra}
  observation,
eV-level spectral resolution observations of the system (as those
that will be carried out by the {\it ASTRO-E2} satellite) can be
used to constrain the dynamics of the disk and establish whether the
accretion disk of Cygnus X-1 actually precesses, and if so, to
determine its dynamics.

\section{Disk precession in Cygnus X-1}

Several X-ray binaries present periodic behavior in their light
curves on timescales longer than the orbital period. Among these
systems we can mention Her X-1, SS 433, and LMC X-4. It has been
suggested that these long periods correspond to the precession of
the accretion disk (e.g. Katz 1973). In the case of SS 433 the
precession is directly measured in the jets, so if these are
attached to the accretion disk it is reasonable to expect that the
disk will also display precession (Katz 1980). Although there is
  no reported compelling evidence yet for disk precession in black hole
  binaries, it is reasonable that the same mechanisms responsible for
  this phenomenon in neutron binaries will apply.

The mechanism that produces the precession might be the instability
of the response of the disc to the radiation reaction force from the
illumination by the central source (e.g, Wijers \& Pringle 1999,
Ogilvie \& Dubus 2001), or the tidal force of the companion star on
a disk which is not coplanar with the binary orbit (Katz 1973,
Larwood 1998, Kaufman Bernad\'o et al. 2002). Uniform disk
 precession will occur in this case only if the sound crossing time
through the disk is considerably shorter than the characteristic
precession period induced by the perturbing star.
%
%
The precession angular velocity is given by (e.g. Romero et al.
2000): $ \left|\Omega_{\rm p}\right|\approx \frac{3}{4}
\frac{Gm}{r_{\rm m}^3} \frac{1}{\omega_{\rm d}} \cos\theta, $ where
$G$ is the gravitational constant, $r_{\rm m}$ is the orbital
radius, $\omega_{\rm d}$ is the inner disk angular velocity,
$\theta$ is the half-opening angle of the precession cone, and $m$
is the mass of the star that exerts the torque upon the disk. The
orbital period $T_{\rm m}$ is related with the involved masses and
the size of the orbit by Kepler's law: $ r_{\rm
m}^3=\frac{G(m+M)T_{\rm m}^2}{4\pi^2}, $ where $M$ is the mass of
the accreting object. The ratio between the orbital and the
precessing periods can be related through the disk angular velocity
$\omega_{\rm d}=(GM/r_{\rm d}^3)^{1/2}$:
\begin{equation}
\frac{T_{\rm m}}{T_{\rm p}}=\frac{3}{4}\frac{m}{M} \kappa^{3/2}
\left(\frac{M}{m+M}\right)^{1/2} \cos \theta, \label{ratio}
\end{equation}
where $T_{\rm p}=2\pi/\Omega_{\rm p}$ and $\kappa=r_{\rm d}/r_{\rm
m}$. Since $\kappa<1$, normally $T_{\rm m}/T_{\rm p}<1$.
In the case of Cygnus X-1, Brocksopp et al. (1999) have reported
multiwavelength evidence for the existence of a period of
$142.0\pm7.1$ days. Similar precessing periods have been calculated
by Larwood (1998), Katz (1973, 1980) and Katz et al. (1982) for
other X-ray binaries on the basis of the same model. As shown by
Romero et al. (2002) in the case of Cygnus X-1, such a period can be
obtained from tidally-induced precession for an accretion disk with
a size $\sim 4\times 10^{11}$ cm, if the half opening angle of the
precession cone is $\sim 15^{\circ}$. For a purely wind-fed system,
this size might be too large and other mechanism may be in operation
to generate
%
%
the observed timescales. In particular, radiation-driven precession
(Pringle 1996, Maloney \& Begelman 1997, Ogilvie \& Dubus 2001),
wind-driven warping and precession (Schandl \& Meyer 1994, Quillen
2001), and spin-spin precession (Bardeen \& Petterson 1975, Armitage
\& Natarajan 1999) can yield precession periods of several weeks to
a few months. In all these mechanisms the observed optical and X-ray
modulation points to precession of the disk, whereas the radio
variations might originate in the jet.


\section{Fe K-$\alpha$  line profile diagnosis}

The use of emission lines as a diagnosis tool for the state of
binary systems has been proposed in the past (e.g., for an
investigation on supermassive black hole binarity, see Torres et al.
2003, also Gaskell 2003; Zakharov et al. 2004a,b). Here we show that
a similar method can be used to extract information about the
precession status of microquasars.

For the case of Cygnus X-1, we shall assume that the time-averaged
disk inclination angle is 35$^\circ$, which is in agreement with the
fitting of the system's Fe line (Miller et al. 2002). Values around
this time-averaged inclination angle were also found for other
binary systems ( e.g., Her X-1, LMC X-4, SMC X-1, etc. e.g., as
discussed by Larwood 1998). We also assume two extreme cases for the
amplitude of the disk precession: (1) the disk inclination angle
precesses from 31$^\circ$ to 39$^\circ$ (very low magnitude of the
precession angle); (2) the disk inclination angle precesses from
5$^\circ$ to 65$^\circ$ (large magnitude of the precession angle).
The amplitude of the precession of the inner disk should not be
large if it is due to the tidal force of the secondary star and if
the disk (especially the outer disk region) develops a significant
warp. However, if the initial spin direction of the BH is
significantly different from the orbital angular momentum direction,
the inner disk, which is confined to the equatorial plane of the BH
(if the spin is high) due to the Bardeen-Peterson effect, may
precess around the total angular momentum (dominated by the orbital
angular momentum) with an amplitude as large as the initial orbital
inclination angle with respect to the BH equatorial plane. If we
assume that the disk is rigidly precessing around the total angular
momentum (dominated by the orbital angular momentum), then the
amplitude of the precession may also be around 30$^\circ$.


Several calculations on the disk line profiles have been performed.
We use a ray-tracing technique and elliptic integrals (Rauch \&
Blandford 1994; see also  Yu \& Lu 2000; Lu \& Yu 2001 and
references therein) to follow the trajectories of photons from the
observer, keeping track of all coordinates until the photons either
intersect the accretion disk plane, disappear below the event
horizon, or escape to "infinity"  (operationally defined to be
$r=1000GM/c^2$ away from the BH). We then calculate the redshift
factor for a photon (to the observer) emitted from a particular
position on the disk. The solid angle subtended at the observer by
each disk element is also calculated.  We set the inner radius of
the disk to be at the marginally stable orbit ($6r_{\rm g}$ for a
Schwarzschild black hole or $1.23r_{\rm g}$ for a Kerr black hole
with spin $a/M=0.998$, where $r_{\rm g}=GM/c^2$), and the outer
radius at $160r_{\rm g}$. We assume that the surface emissivity of
line photons follows a power-law, $r^{-q}$, with $q=2.5$. Both the
power-law emissivity law and the size of the disk in Schwarszchild
units, are usual assumptions (see, e.g., Nandra et al. 1997). The BH
spin is assumed to be $a/M=0.998$. In microquasar systems, both the
high frequency quasi-periodic oscillation and relativistic lines
suggest a high spin. In any case, we proved that if we were to
assume a lower spin, there is not much qualitative difference for
the problem we have studied here. With the above assumptions, we sum
up all the photons received by the observer, which is emitted from
each disk element, and obtain the profile of emergent Fe K$\alpha$
lines, with different inclination angles from the Cygnus X-1 system,
as shown in the Figure 1.


\section{Observing the Fe line profile variations}

The complex X-ray spectrum of Cygnus X-1 revealed by the {\it
Chandra} observation reported in Miller et al (2002) implies that
the detection of minute variations in the Fe line shape will require
a high spectral resolution instrument with large throughput.  In
what follows, we discuss the feasibility of detecting the precession
of the accretion disk with {\it ASTRO-E2}, and specifically with the
X-ray Spectrometer (XRS) consisting of an array of
semiconductor-based calorimeters delivering the best spectral
resolution to date at 6 keV (pre-flight value of 6.5 eV). {\it
ASTRO-E2} is a JAXA/NASA mission to observe X-rays with
unprecedented high spectral resolution imaging detectors, which is
scheduled for launch by Summer 2005. The {\it ASTRO-E2} Science
Working Group target list includes two observations of Cygnus X-1.

The underlying assumption is that the X-ray spectrum of Cyg X-1
  can be well described by the best-fit model obtained by Miller et al
  (2002). Indeed, at higher spectral resolution, other -so far
  undetected- spectral components might appear.  Examples include the
  presence of blue-shifted edges due to high velocity ejecta. Should
  these, or other, features impact strongly either on the Fe line
  region itself or in the energy ranges where the underlying continuum
  is estimated, the specific methodology proposed here will need to be
  revised. Fortunately the instrument selected will deliver spectra
  with a very high signal to noise at a spectral resolution so high
  that many of these putative components might be resolved and modeled
  out.  That would certainly complicate, but not invalidate, the
  proposed analysis.

As argued by Miller et al (2002) the spectrum of Cygnus X-1 is
complex below 3 keV, where they did not succeed in fitting an
appropriate model. All our discussion assumes the Be filter is on at
the XRS. This removes low energy photons which are not needed for
our purposes. In our analysis,  we ignore photons below 3 keV and
above 9 keV. 
In addition, we do not use
the X-ray Imaging Spectrometers (XIS) on board {\it ASTRO-E2}, which
have less spectral resolution at the Fe line energy range. A typical
50 ks exposure is assumed, and the background is assumed to be
negligible for such an strong source (see below) and not included in
the simulations. Pre-launch calibration redistribution matrices and
efficiency curves have been downloaded from the {\it ASTRO-E2} web
site at
NASA\footnote{http://astroe2.gsfc.nasa.gov/docs/astroe/prop\_tools/xrs\_mat.html,
  with the most recent updates included (as of August 4, 2004) }.

The simulated model is that fitted by Miller et al. (2002) to the
{\it Chandra} data. The continuum is a power law with $\Gamma=1.8$,
absorbed by a Galactic column of $N_H=8.1\times 10^{21}\, {\rm
cm}^{-2}$
%
%
which is absorbed by a smeared edge at 7.2~keV, with a width of 7
keV and a depth of 1.2. The narrow line component at 6.415 keV,
believed to arise from the irradiated outer disk, has been simulated
as a gaussian of width $\sigma_{\rm narrow}=30$ eV. The broad line
component has been simulated using the numerical models explained in
the previous section, for a fixed equivalent width of 140 eV and a
variety of disk inclination angles.

As expected, this model produces a very high count rate in the XRS,
of the order of 60 counts s$^{-1}$. This count rate will be
distributed among several XRS pixels, according to the PSF. Although
the overall count rate is below the telemetry limit, the fraction of
events that will be measured by the on-board software as medium or
low resolution will be large ($\sim 30-40\%$). There are two
possibilities to deal with this: either using the neutral density
filter (which will decrease the overall count rate to an acceptable
level of $\sim 6$ counts s$^{-1}$  ) or to ignore the few pixels
with the higher count rates and work only with the pixels which have
count rates below a few counts s$^{-1}$. Both of these procedures
will result in an overall loss of throughput. This is why we
consider {\it effective exposures} of 10 and 5 ks.

To analyze the simulated data, we follow Miller et al (2002) to fit
the continuum, by excluding the range from 4.0 to 7.2 keV.  A single
power law leaves enormous residuals which can be well fitted by the
smeared absorption edge.  In general, the edge energy is well
reproduced by the fit (statistical 90\% errors in the range of
50--150 eV depending on the effective exposure time), although there
is substantial degeneracy between the width of the edge and its
depth. This does not affect the continuum in the Fe emission line
region.

Once the continuum is fitted, we include all data in the 3.0-9.0
energy band, and add a narrow gaussian emission line and a
relativistic emission line (see Figure 2, left).  Thanks to the
superb spectral resolution of this instrument, the narrow line is
very well characterized, with errors in its centroid actually
limited by systematics (2 eV) rather than by statistics, even in a 5
ks exposure. The relativistic line model returns the disk
inclination angle with a 90\% error of 0.3 deg for a 50 ks exposure
and 0.7 deg for a 5 ks exposure.  This is due to the fact that the
sharp drop in the blue edge of the line is very clearly marked by
the XRS. Figure 2 (right panel) shows the differences in these sharp
edges for a 50 ks exposure and various disk inclination angles. Note
that these changes amount to about 50 eV per degree of disk
inclination, and therefore the claimed limit in the systematics for
line centering of 2 eV is really not an issue for this purpose.
Table 1 summarizes the results of the fits to simulated data with 5,
10 and 50 ks net exposure and disk inclination of $35^{\circ}$.

\section{Concluding remarks}

This work suggests a diagnostic tool to investigate whether the
accretion disk of Cygnus X-1 (and of other microquasar systems)
precesses, and if so, what is the period and precession angle. We
have shown that the study of the periodic variations of the Fe
K$\alpha$ line, that would be unavoidably produced in the putatively
precessing disk of the system, are observable in short (5--10 ks)
exposures of the {\it ASTRO-E2} satellite. The degree of precision
and confidence level up to which we will be able to determine the
inclination angle of disk for each short observations, thus, the
magnitude of the precession, was shown to be sufficiently high as to
allow a clear determination of these parameters even when the
precession angle is of only a few degrees.

\acknowledgements

The  work of DFT was performed under the auspices of the U.S.
DOE--NNSA by U. of California LLNL under contract No. W-7405-Eng-48.
GER is supported by the research grant PICT 03-13291 (ANPCT).  XB
acknowledges financial support by the Spanish Ministerio de
Educaci\'{o}n y Ciencia, under project ESP 2003-00852.

\clearpage

\begin{figure*}[t]
\centering
\includegraphics[width=6cm,height=7cm]{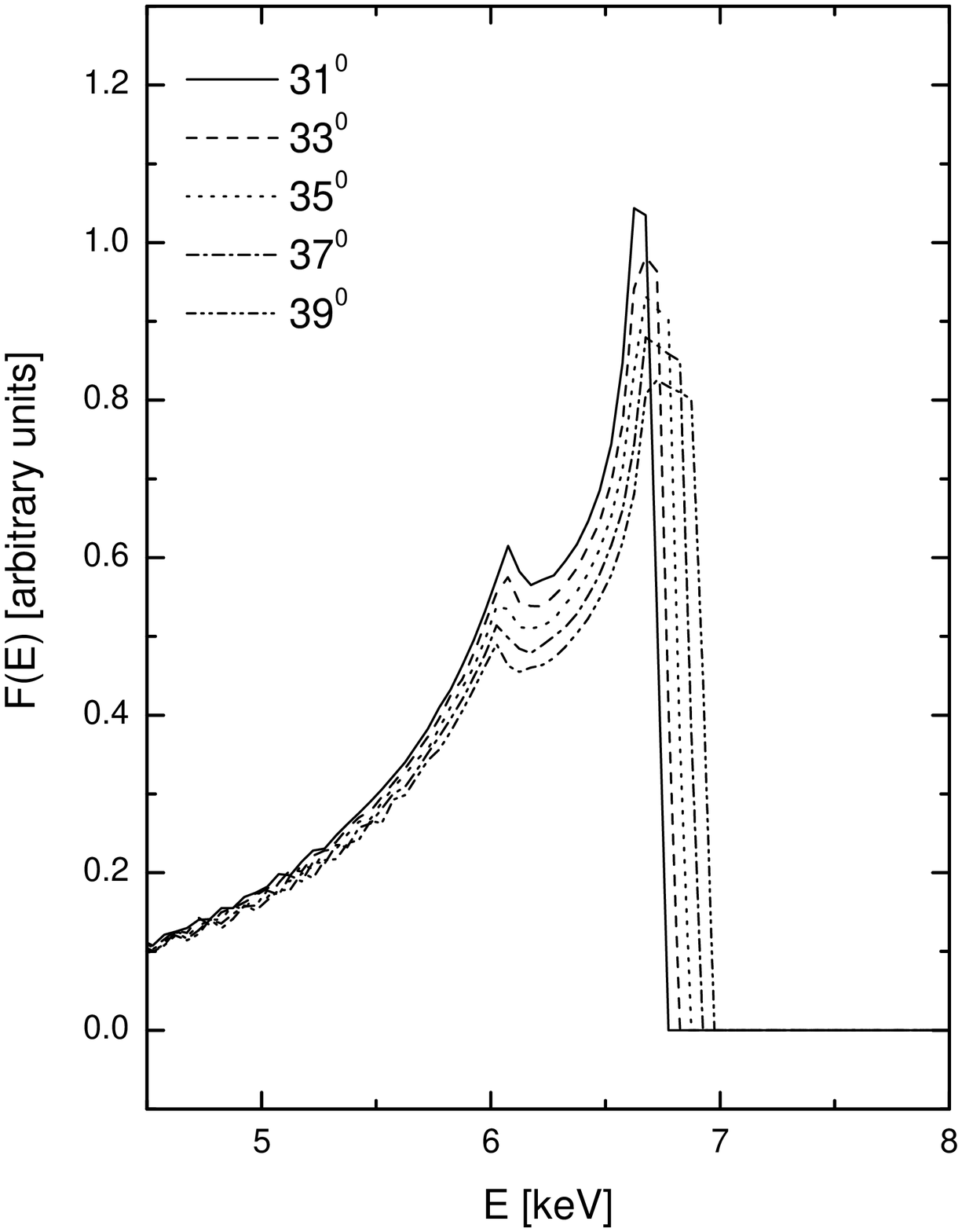}
\includegraphics[width=6cm,height=7cm]{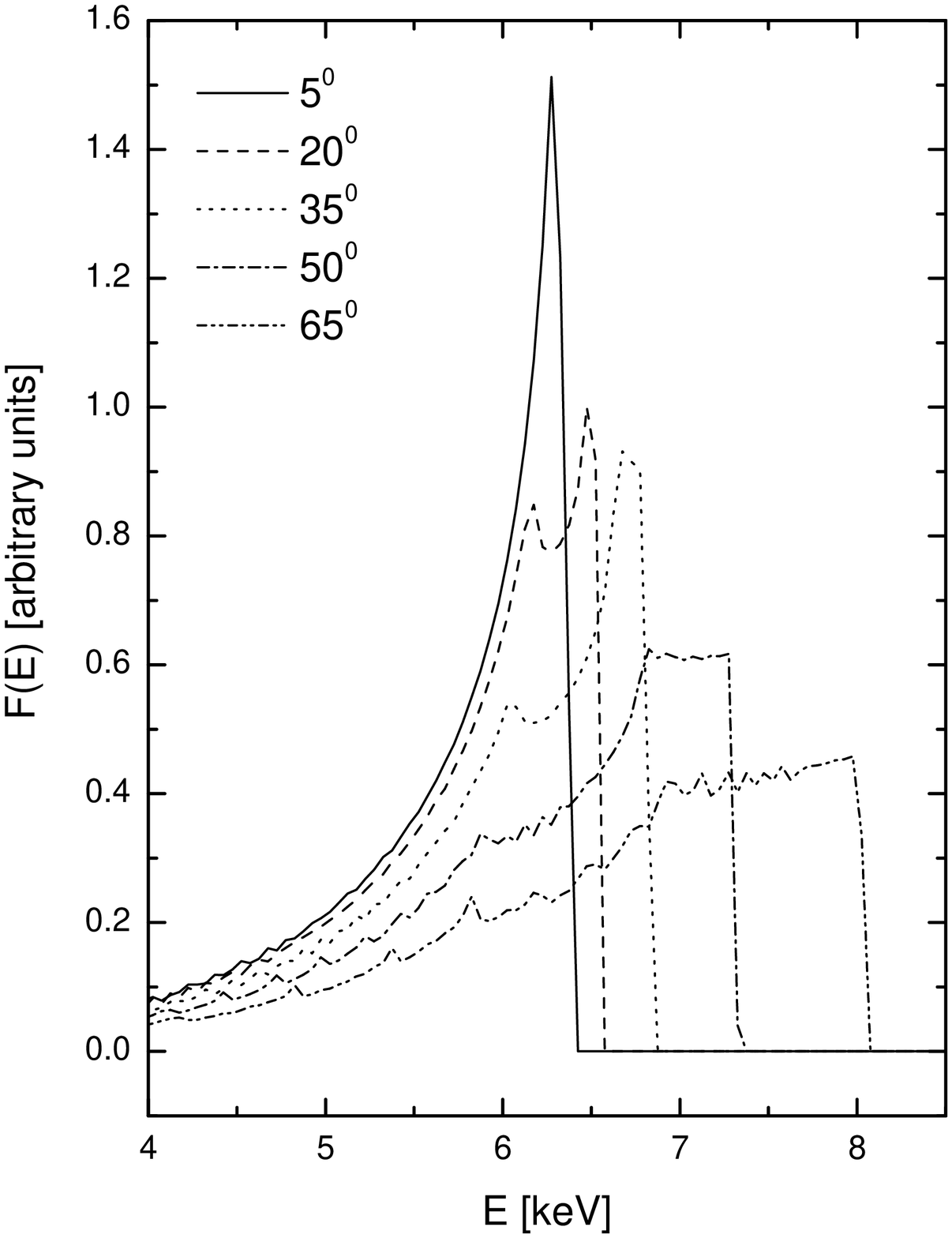}
\caption{Shift in the Fe  K-$\alpha$ line profile of Cyg X-1 as a
function of different inclination.
The total line
flux of each line is normalized to 1. }
\end{figure*}

\clearpage

\begin{figure*}
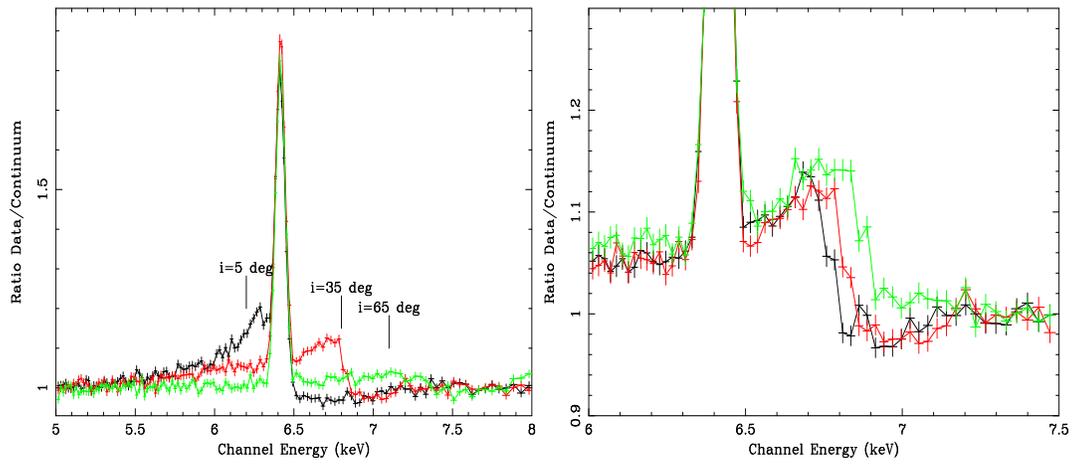

\centering \label{ratios1}
\includegraphics[width=6cm,height=7cm,angle=-90]{f2a.ps}
\includegraphics[width=6cm,height=7cm,angle=-90]{f2b.ps}
\caption{Left: Ratio of data to fitted continuum for net 50 ks
exposures, assuming inner disk inclination angles of $5^{\circ}$,
  $35^{\circ}$ and $65^{\circ}$. Right: Detail
of the ratio of data to fitted continuum for net 50 ks
  exposures, assuming inner disk inclination angles of $33^{\circ}$,
  $35^{\circ}$ and $37^{\circ}$ with the drop of the broad line going
  from left to right. Data have been grouped for fitting and presentation
purposes.}
\end{figure*}

\clearpage

\begin{table*}
\begin{center} \label{fits}
\centering \caption{Disk parameters, with 90\% errors,
for various effective exposure times}
\begin{tabular}{l c c c c}
\hline
Parameter & input & 50 ks & 10 ks & 5 ks\\
\hline
$\Gamma$ & $1.80$ & $1.80\pm 0.010$ & $1.80\pm 0.015$ & $1.80\pm 0.02$\\
$E_{\rm edge}$ (keV) & $7.20$ & $7.19\pm 0.05$ & $7.20\pm 0.10$ & $7.12\pm 0.15$\\
$E_{\rm narrow}$ (keV) & $6.415$ & $6.414\pm 0.001$ & $6.414\pm
0.001$ &
$6.416\pm 0.002$ \\
$\sigma_{\rm narrow}$ (eV) & $30$ & $29.3\pm 0.6$ & $29.3\pm 1.2$ &
$32.0\pm 2.0$ \\
Disk inclination & $35$ & $35.0\pm 0.3$ & $35.2\pm 0.6$ & $35.3\pm
0.7$\\
\hline
\end{tabular}
\end{center}
\end{table*}


\begin{thebibliography}{}



\bibitem[]{428} Armitage, P.J. \& Natrarajan, P. 1999, ApJ, 525, 909

\bibitem[]{430} Bardeen, J.M. \& Petterson, J.A. 1975, ApJ, 195, L65

\bibitem[]{454}Barr, P., White, N. E., \& Page, C. G. 1985, MNRAS, 216, 65p

\bibitem[]{456}Brocksopp, C., Fender, R. P., Larimov, V., et al. 1999, MNRAS, 309, 1063


\bibitem[]{460}Dove, J. B., Wilms, J., Maisack, M., \& Begelman, M. G. 1997, ApJ, 487, 759



\bibitem[]{467}Ebisawa, K. et al. 1996, ApJ, 467, 419

\bibitem[]{447} Esin A. A., McClintock, J. E. \& Narayan, R., 1997, ApJ, 489, 865,
\bibitem[]{448} Esin A. A., Narayan, R., Cui, W., Grove, J. E. \& Zhang, S. 1998, ApJ, 505, 854

\bibitem[]{469}Fender, R. P. 2001, MNRAS, 322, 31

\bibitem[]{471} Gaskell, C. M. 2003, "Quasars as Supermassive Binaries", Proceedings of
24th Liege International Astrophysical Colloquium, p. 473.


\bibitem[]{475}Gierli\'nski, M., Zdziarski, A. A., Poutanen, J., et al. 1999, MNRAS 309, 496

\bibitem[]{477}Herrero, A., Kudritzki, R. P., Gabler, R., et al. 1995, A\&A 297, 556


\bibitem[]{481} Katz, J.I. 1973, Nature Phys. Sci., 246, 87

\bibitem[]{483}Katz, J. I. 1980, ApJ, 236, L127

\bibitem[]{466} Katz, J.I., Anderson, S.F., Margon, B., \& Grandi, S.A. 1982, ApJ,
260, 780

\bibitem[]{485}Kaufman Bernad\'o, M. M., Romero, G. E., \& Mirabel, I. F. 2002, A\&A, 385, L10

\bibitem[]{487}Kitamoto, S., et al. 1990, PASJ, 36, 731

\bibitem[]{489} Lu, Y., \& Yu, Q. 2001, ApJ, 561, 660



\bibitem[]{494} Larwood, J.D. 1998, MNRAS, 299, L32

\bibitem[]{480} Maloney, P.R. \&
Begelman, M.C. 1997, ApJ, 491, L43

\bibitem[]{496}Miller, J. M. et al. 2002, ApJ, 578, 348

\bibitem[]{498} Nandra, K., George, I. M., Mushotzky, R. F., Turner, T. J., \&
Yaqoob, T., 1997, ApJ, 477, 602


\bibitem[]{504} Ogilvie G. I. \& Dubus G. 2001, MNRAS 320, 485




\bibitem[]{508}Poutanen, J., Krolik, J. H., \& Ryde, F. 1997, MNRAS, 292, L21

\bibitem[]{499} Pringle, J.E. 1996, MNRAS, 281, 357

\bibitem[]{501} Quillen, A.C.
2001, ApJ, 563, 313

\bibitem[]{511} Rauch, K., \& Blandford, R.\ D.\ 1994, ApJ, 421, 46

\bibitem[]{506} Reynolds, C.S. \& Nowak, M.A., \ 2003, Phys. Rep., 377, 389

\bibitem[]{513} Romero, G. E., Chajet, L., Abraham, Z, Fan, J. H. 2000, A\&A, 360,
57

\bibitem[]{516}Romero, G. E., Kaufman Bernad\'o, M. M., \& Mirabel, I.F. 2002, A\&A, 393, L61

\bibitem[]{513} Schandl, S. \& Meyer, F. 1994, A\&A 289, 149

\bibitem[]{518}Stirling, A. M., Spencer, R. E., de la Force, C. J., et al. 2001, MNRAS, 327, 1273

\bibitem[]{520}Tanaka, Y. 1991, Fe Line Diagnostics in X-Ray Sources, ed. A. Treves (Berlin, Springer), 98

\bibitem{Torres:2003cc}
Torres, D. F., Romero, G. E., Barcons, X. \& Lu, Y. J. 2003, ApJ
596, L31


\bibitem[]{532} Wijers, R. A. M. J. \& Pringle J. E. 1999, MNRAS 308, 207


\bibitem[]{536} Yu, Q., \& Lu, Y., 2000, MNRAS, 311, 161


\bibitem[]{539} Zakharov, A. F., Ma, Z. \& Bao, Y. 2004a, New Astronomy 9, 663

\bibitem[]{541} Zakharov, A. F., 2004b, astro-ph/0411611

\end{thebibliography}
\end{document}